\documentclass[proof]{WileyASNA-v1}

\articletype{PROCEEDING}%

\received{XXX}
\revised{XXX}
\accepted{XXX}

\newcommand{\integral}{\textit{\integral}\xspace}

\begin{document}

\title{INTEGRAL observations of magnetars\protect}

\author[1,2]{Dominik Patryk Pacholski*}

\author[3,4]{Lorenzo Ducci}

\author[5]{Martin Topinka}

\author[1]{Sandro Mereghetti}

\address[1]{\orgdiv{INAF}, \orgname{ IASF-Milano}, \orgaddress{via A. Corti 12, 20133 Milano, \country{Italy}}}

\address[2]{\orgdiv{Universit\`a degli Studi di Milano Bicocca}, \orgname{Dipartimento di Fisica G. Occhialini}, \orgaddress{Piazza della Scienza 3, 20126 Milano, \country{Italy}}}

\address[3]{\orgdiv{Institut f\"ur Astronomie und Astrophysik},  \orgaddress{\state{T\"ubingen}, \country{Germany}}}

\address[4]{\orgdiv{ISDC}, \orgname{University of Geneva},  \country{Switzerland}}

\address[5]{\orgdiv{INAF}, \orgname{Osservatorio Astronomico di Cagliari},   \country{Italy}}

\corres{*Corresponding author  \email{dominik.pacholski@inaf.it}}

\abstract{The INTEGRAL satellite has collected a large amount of data on magnetars in our Galaxy, spanning more than 20 years starting from 2003. 
The large data set obtained with the IBIS/ISGRI instrument at energies above 20~keV allows us to study both the properties and long-term evolution of their persistent hard X-ray emission and the population characteristics of the short bursts emitted during active periods. 
We are carrying out a comprehensive analysis of the observed magnetars, exploiting the most recent calibrations and analysis software. Here we report on the long term evolution of the hard X-ray flux of the magnetars detected with ISGRI and  the results of a sensitive search for short bursts in SGR J1935+2154.}


\maketitle

\section{Introduction}\label{sec1}

Magnetars are a small class of young neutron stars, with only around 30 known sources \citep{2014ApJS..212....6O}. They are powered by their strong magnetic field, which can reach up to $10^{15}$~G \cite[e.g.,][]{2015SSRv..191..315M,2017ARA&A..55..261K}. They exhibit a variety of emissions, ranging from short bursts that last only a few tens of milliseconds   to long outbursts that can persist for weeks or even months. During these outbursts, the persistent emission can increase by 10 to $10^{3}$ times above its normal level \cite[e.g.,][]{2018MNRAS.474..961C}.  The outbursts are often correlated with an increased number of bursts, and are followed by long periods of quiescence, during which the emission returns to a quiescent level of $10^{30}-10^{33}$~erg~s$^{-1}$.
 In the hard X-ray range (>10~keV), a few magnetars display a  hard power-law tail extending up to $\sim200$~keV, with emission energetically comparable to the one in the soft energy range  \citep{2005A&A...433L..13M,2006A&A...451..587D,2006ApJ...645..556K,2007A&A...475..317G,2015A&A...583A.113D}.

INTEGRAL is a  satellite of the European Space Agency operating since the end of 2002. It has extensively observed the Galactic plane, thus providing long-term coverage of the whole population of Galactic magnetars.
Here we concentrate on data obtained with  IBIS (Imager on Board of the INTEGRAL Satellite,   \cite{2003A&A...411L.131U}), which is a  coded-mask instrument comprising  two detectors, ISGRI and PICsIT. ISGRI operates in the $\sim20-1000$~keV range, providing photon-by-photon data that allow high-resolution timing and imaging with 12 arcmin angular resolution in a field of view of $~29\times29$ deg$^2$ \citep{2003A&A...411L.141L}.

We used data of the ISGRI detector to study both the persistent hard X-ray emission and the short bursts of magnetars, exploiting the most recent software, Off-line
 Scientific Analysis (OSA) version 11.2 \citep{2003A&A...411L.223G},  and calibrations.

\section{Data analysis and results}

\subsection{Long term light curves}

In order to optimize the imaging performances and reduce systematic background effects, INTEGRAL observes using a  dithering  strategy. As a result, the data are split into short  pointings called Science Windows (ScWs) with typical duration of $\sim$30 min each.
 
\begin{figure*}[t]
\centerline{\includegraphics[width=180mm,height=22.5pc]{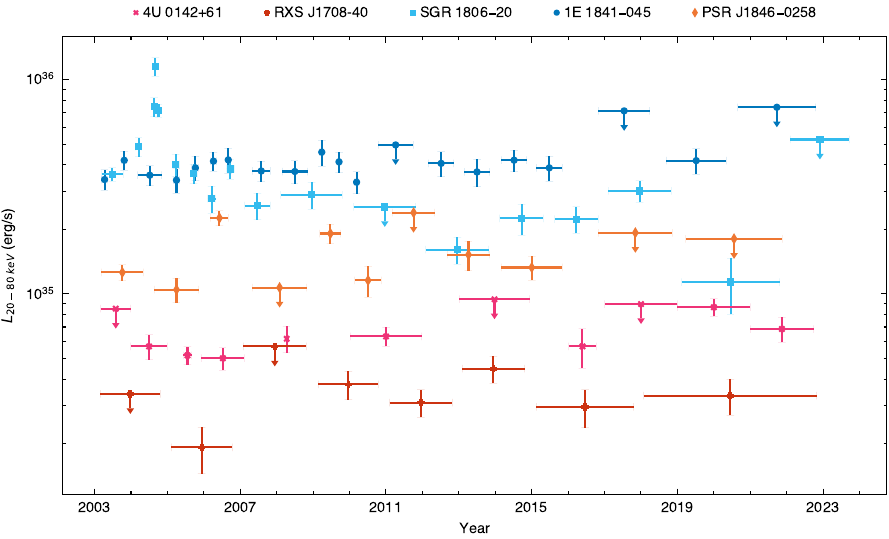}}
\caption{Long term evolution of the luminosity in the 20-80~keV energy range for five magnetars. The adopted distances are 3.6 kpc for 4U 0142+61, 3.8 kpc for RXS J1708-40, 8.7 kpc for SGR 1806-20, 8.5 kpc for 1E 1841-045, and  6.0 kpc for PSR J1846-0258. \label{fig:fig1}}
\end{figure*}

To study the long term evolution of the hard X-ray emission, we selected all the ScWs pointed within $10^{\circ}$ from the source of interest. This guarantees a good sensitivity and a well known response of the instrument. We removed all the ScWs affected by high background or other instrumental  problems.

The lower energy threshold of the ISGRI instrument evolved during the mission due to the ageing of the detector, from $\sim$18~keV in the early years to the current value of $\sim$30~keV. In order account for this effect  without loosing too much sensitivity,  we created images in the broad energy range [$E_{min}-80$~keV],  with $E_{min}$ calculated to ensure stable response of the instrument and taking progressively increasing values from 18 to 30~keV as the mission evolved\footnote{see section Known Limitations in the IBIS Analysis User Manual available at the ISDC website \url{http://isdc.unige.ch/integral/analysis}}.

In general, the persistent emission of our targets is too faint to be detected in single ScWs.  Therefore, we stacked ScWs for time intervals sufficiently long to reach     a detection significance of at least $7\sigma$ for each magnetar.   This value was chosen after several   tests which showed   that lower confidence levels may lead to an overestimate of the  flux.
 
We then used the \texttt{mosaic\_spec} tool in OSA to extract from the created mosaics the background-subtracted  count rates of the source. A 5\% systematic uncertainty was added to the count rate statistical errors.
These count rates (extracted in different energy intervals) were then converted to fluxes in the 20-80~keV range, using the   conversion factors derived from the proper (time-dependent) response files created \textit{ad hoc} for each mosaic. 

Since the magnetar hard X-ray emission in the analysed energy range is usually well described by a  power law with photon index $\Gamma$ between 1 and 2, we assumed $\Gamma=1.7$ and, to account for possible  different slopes, we included an additional   5\% flux uncertainty,  that would result for $\Gamma$ between 1 and 2.5.

We     could    create long term lightcurves (shown in Figure \ref{fig:fig1}) for the  five sources listed  in Table \ref{tab:tab1}.
We also include the allegedly rotation-powered PSR J1846--0258 which exhibited magnetar-like behavior in 2006 \citep{2008Sci...319.1802G,2009A&A...501.1031K} and in 2020 \citep{2020ATel13985....1K}.
Three more magnetars were detected  only for short time  periods and are not shown in the figure. They are SGR 0501+4516	and 1E 1547.0--5408, detected during an outburst occurring in 2008 and 2009, respectively, and SGR 1900+14, which was visible in 2004 and  had only a few low-confidence detections in the following years.
The variability factor, defined as  $V_{f} = F_{max}/F_{min}$,  and the error weighted average flux of the sources are listed in Table \ref{tab:tab1}.

\begin{center}
\begin{table}[t]%
\centering
\caption{List of detected magnetars with their average X-ray fluxes ($F_{mean}$) and variability factors ($V_f$).
\label{tab:tab1}}%
\tabcolsep=0pt%
\begin{tabular*}{20pc}{@{\extracolsep\fill}lcccc@{\extracolsep\fill}}
\toprule
\textbf{Source} & \textbf{F$_{mean}^{a}$} & \textbf{V$_f$} \\
\midrule
4U 0142+61 & $3.83 \pm 0.16$ & $1.9\pm0.4$ \\
RXS J1708--40 & $1.80 \pm 0.13$ & $3.0\pm0.8$ \\
SGR 1806--20 & $3.31 \pm 0.10$ & $10.2\pm3.1$ \\
1E 1841--045 & $4.42 \pm 0.12$ & $2.2\pm0.4$ \\
PSR J1846--0258 & $3.26 \pm 0.14$ & $2.3\pm0.4$ \\

\bottomrule
\end{tabular*}
\begin{tablenotes}
\item $^{a}$ Error weighted average flux in the $20-80$~keV energy range in units of 10$^{-11}$~erg~cm$^{-2}$~s$^{-1}$ excluding upper limits.
\end{tablenotes}
\end{table}
\end{center}

\subsection{Bursting activity from SGR J1935+2154}

SGR J1935+2154 has been one of the most active magnetars in recent years. Since its discovery in 2015, it went through multiple outbursts \citep{2017ApJ...847...85Y,2020ApJ...904L..21Y,2022MNRAS.516..602B,2020ApJ...893..156L,2020ApJ...902L..43L,2024ApJ...965...87I} connected with bursting activity. During one of the most active periods, in April 2020, it emitted a so-called ``burst forest'' consisting of more than 200 burst within 20 minutes \citep{2020ApJ...904L..21Y,2021ApJ...916L...7K,2022ApJS..260...24C}. A few hours later it also provided evidence for a connection between magnetars with Fast Radio Burst (FRB), because it emitted a strong FRB-like burst of radio emission simultaneously  with a hard spectrum X-ray burst \citep{2020Natur.587...59B,2020Natur.587...54C,2020ApJ...898L..29M,2021NatAs...5..378L,2021NatAs...5..401T,2021NatAs...5..372R}.
We searched for all public ScWs in the INTEGRAL archive in which source was within $14.5^{\circ}$ from the pointing direction. We eliminated the ScWs affected by strong and/or highly variable background, finally resulting in  12409 ScWs for a total observing time of about 31~Ms.

For each ScW, we selected only events registered by the detector's pixels that were illuminated by the source for at least 50\% of their surface. We extracted   15-300~keV   light curves in seven logarithmically-spaced time bins of duration between 10~ms and 1.28~s. These were then examined looking for bins in which counts significantly exceeded a threshold corresponding to a 3$\sigma$ level, accounting for the total number of bins. 
All the triggers  found in the light curves with this procedure were then examined exploiting the imaging capabilities of the detector to confirm their nature and association with the magnetar.

This finally led to a sample of 182 bursts, most of which (149) were emitted during the outburst in October 2022. In particular, we observed 137 bursts between October 12 at 12:47 UT and October 13 at 00:56 UT. Two minutes after the last burst, there was a gap in observation until October 14 at 6:43 UT. The next burst was observed at 20:48:37 UT, after more than 38 ks of observation time. In the following analysis, we focused on the 149 bursts of the October 2022 outburst. 

\begin{figure}[t]
	\centerline{\includegraphics[width=78mm,height=13.5pc]{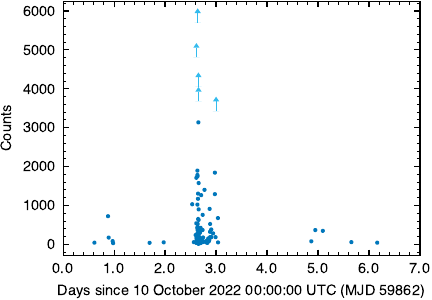}}
	\caption{Corrected number of counts for each burst detected by ISGRI in the October 2022 outburst of SGR J1935+2154. Saturated bursts are indicated by light-blue arrows representing the lower limit.\label{fig:cts}}
\end{figure}

To derive the burst properties, we  corrected the number of detected counts to  take into account the off-axis position and the dead-time of the detector, which is equal to $~114\mu$s for each illuminated module of the detector \citep{2006A&A...445..313G}.
In Figure \ref{fig:cts} we show the resulting number of counts for each burst as a function of time.

\begin{figure*}[t]
\centerline{\includegraphics[width=180mm,height=13.5pc]{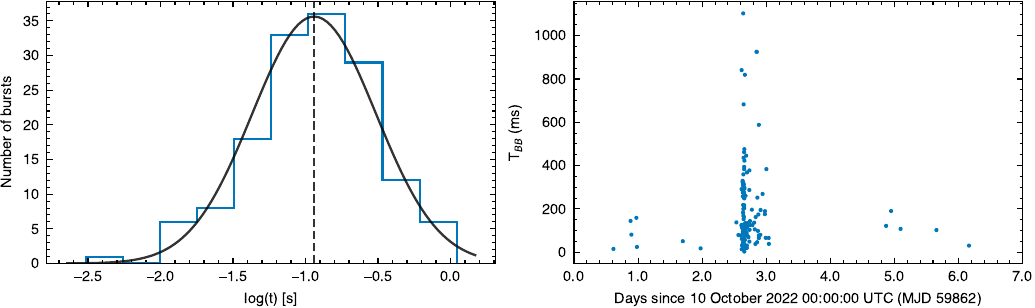}}
\caption{T$_{BB}$ distribution of the October 2022 bursts of SGR J1935+2154 (left panel) and distribution of T$_{BB}$ over MJD time (right panel). \label{fig:tbb}}
\end{figure*}

We estimated the duration $T_{BB}$ of the bursts using Bayesian Blocks \citep{1998ApJ...504..405S}. This technique segments a non-binned event list into time intervals, each of which has no statistically significant variability within itself. This allows the identification of significant change points that mark the start of new intervals, enabling a more precise determination of the burst duration.

We plotted in   Figure \ref{fig:tbb} (left panel) the distribution of the logarithms of the durations. 
It  is well fit by a Gaussian with mean $114\pm9$~ms.
The plot of burst duration over time (right panel of Figure \ref{fig:tbb})  shows that the longest bursts occurred mostly during the peak of bursting activity.

We selected bursts with at least 300 counts to perform spectral analysis.
We defined start and stop times for spectral extraction based on the T$_{BB}$ durations.  Six  bursts were affected   by telemetry saturation  which caused gaps in their lightcurves. For their spectral analysis, we excluded gap periods, and we treat the obtained flux as a lower-limit.

\begin{figure*}[t]
\centerline{\includegraphics[width=180mm,height=13.5pc]{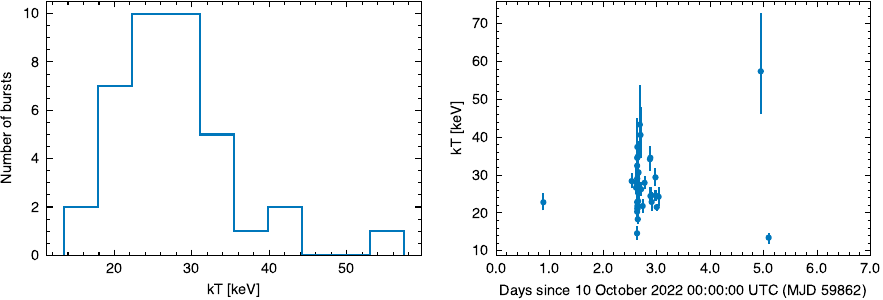}}
\caption{Distribution of the temperatures derived for the bursts with the OTTB model (left panel) and temperatures as a function of time (right panel). \label{fig:kt}}
\end{figure*}

We used OSA to extract spectra of the bursts in 12 logarithmic bins between 30-300~keV. We performed spectral analysis with Xspec version 12.12 \citep{1996ASPC..101...17A}. We used two models: an optically thin thermal bremsstrahlung (OTTB) and an exponentially cut-off power law (COMPT). The single component OTTB model provided a good fit for 38 bursts, yielding an error-weighted average temperature of $kT=23.4\pm0.3$~keV, while the COMPT model had a good fit in case of 16 bursts, with the average parameters $\alpha=-0.03\pm0.21$ and $E_{peak}=35.1\pm1.3$~keV.
In   Figure \ref{fig:kt} we show the   distribution of the OTTB model  temperatures in the left panel and the temperature vales as a function of time in the right panel. 

The brightest burst was detected at 15:20:30.95 UT of 2022-10-12. It had an average flux of at least $(1.46\pm0.05)\times 10^{-6}$~erg~cm$^{-2}$~s$^{-1}$,  and a fluence of $(1.60\pm0.05)\times 10^{-6}$~erg~cm$^{-2}$  (these are lower limits because this burst was affected by telemetry saturation). The burst spectral shape was not  significantly different from that of the other bursts, as the OTTB model gave a temperature of 
$kT=21.2\pm0.8$~keV, and the COMPT model gave $\alpha=0.2\pm0.6$ and $E_{peak}=27.9^{+2.6}_{-4.2}$~keV.

\begin{figure}[t]
	\centerline{\includegraphics[width=78mm,height=13.5pc]{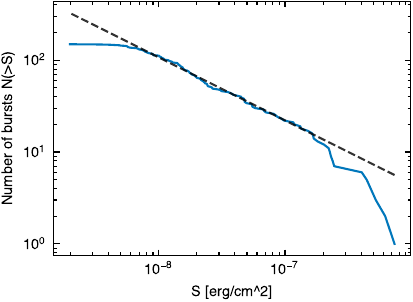}}
	\caption{Cumulative number of bursts with the fluence higher than S. Black line shows fit of power law with the slope of $0.69$ above $6.71\times10^{-9}$~erg~cm$^{-2}$.\label{fig:lognlogs}}
\end{figure}

We calculated the fluences  (30--300 keV) of the bursts based on the detected counts and converted them to physical units using the average parameters found with the OTTB fits. This resulted in an energy to count conversion factor  ECF of $1.4\times10^{-10}$~erg~cm$^{-2}$. The integral distribution of the burst fluences is shown in Figure \ref{fig:lognlogs}. It is well fitted with a power law with index  $-0.69\pm0.01$ for 30-300~keV fluences greater than $6.71\times10^{-9}$~erg~cm$^{-2}$.

\section{Conclusions}

We have reported some results of our ongoing systematic reanalysis of all the magnetars observed with INTEGRAL. 
The long term light curves of the hard X-ray flux spanning more than 20 years show that the most variable source is SGR 1806--20, which emitted a giant flare in December 2004 \citep{2005ApJ...624L.105M}. Its hard X-ray emission peaked in October 2004, before the giant flare, and then it decreased by a factor of 10 in a period of 15 years.  The other sources show only small variability, less than a factor of 3.

SGR J1935+2154 had one of the most active episodes on   12th October 2020. In about 12 hours, INTEGRAL detected 137 bursts. 
For comparison, only 45 other bursts were detected by INTEGRAL in the remaining 31 Ms in which it observed this magnetar.

The burst durations  follow a log-normal distribution, as it has been observed in other   magnetars \citep[e.g.][]{2001ApJ...558..228G,2004ApJ...607..959G,2015ApJS..218...11C}. The average duration $114\pm9$~ms   agrees with that seen in previous bursting periods, which were in the range of $72-182$ ms \citep{2020ApJ...893..156L,2020ApJ...902L..43L}.

The cumulative distribution of the SGR J1935+2154 burst fluences (LogN-LogS) during the October 2022 activity period is well described by a   power law  with   slope of $-0.69\pm0.01$ for 
30-300~keV fluences above  $\sim6.7\times10^{-9}$~erg~cm$^{-2}$.
For our adopted spectral shape, this limit corresponds to $\sim2\times10^{-8}$~erg~cm$^{-2}$
in the 8-200 keV range used by \citet{2020ApJ...893..156L} and \citet{2022ApJS..260...25C} for Fermi/GBM and HXMT data, respectively.
Our LogN-LogS is consistent with that found by these authors. 
\section*{Acknowledgements}
The  results reported in this article are based on data obtained
with INTEGRAL, an ESA mission with instruments and science data centres funded by ESA member states, and with the participation of the Russian Federation and the USA.
This work received financial support from INAF through the Magnetars Large Program Grant (PI S.Mereghetti).

\bibliography{main}%
\end{document}